\documentclass [pra,twocolumn,showpacs]{revtex4}
\usepackage{dcolumn,bm}

\def\etal{{\it et al.}\ }
\newcommand{\eref}[1]{(\ref{#1})}

\def\rtw{\rightarrow}
\begin{document}

\title{Parity non-conservation in thallium}
\author{M. G. Kozlov}
\email{mgk@MF1309.spb.edu}
\author{S. G. Porsev}
\affiliation{Petersburg Nuclear Physics Institute, Gatchina,
188300, Russia}
\author{W. R. Johnson}
\affiliation{Department of Physics, Notre Dame University,
Notre Dame, IN 46556, USA}
\date{\today}

\begin{abstract}
We report a new calculation of the parity non-conserving $E1$ amplitude
for the $6p_{1/2} \rtw 6p_{3/2}$ transition in $^{205}$Tl.
Our result for the reduced matrix element is
$E1_{\rm PNC}=-(66.7\pm 1.7)\,\cdot {\rm i}
\,10^{-11}(-Q_W/N)$~a.u.. Comparison with the experiment of
Vetter \etal  [Phys.\ Rev.\ Letts.\ \textbf{74}, 2658 (1995)] gives the
following result for the weak charge of $^{205}$Tl: $Q_W/Q_W^{\rm SM}
= 0.97\, (\pm 0.01)_{\rm expt}\, (\pm 0.03)_{\rm theor}$, where
$Q_W^{\rm SM}=-116.7\pm 0.1$ is the standard model prediction. This result
confirms an earlier conclusion based on the analysis of a Cs experiment
that atomic PNC experiments are in agreement with the standard
model.
\end{abstract}
\maketitle

\section{Introduction}
\label{intr}

The thallium atom is the second simplest atom where parity
non-conservation (PNC) has been observed \cite{EPBN95,VMM95}. The
simplest and best
understood such atom is cesium, where theory has the accuracy of 1\%
\cite{DFS89a,BSJ92} (see also more recent calculations
\cite{Der00,DHJS01,KPT01}). That, together with the best
experimental result for Cs \cite{WBC97} allowed one to obtain the most
accurate result for the weak charge of the cesium nucleus
\cite{DHJS01,KPT01}:
\begin{eqnarray}
        &&Q_W(^{133}{\rm Cs})=-72.5\,(3)_{\rm expt}(7)_{\rm theor},
\label{i1} \\
        &&Q_W^{\rm SM}(^{133}{\rm Cs})=-73.09\,(3),
\label{i2}
\end{eqnarray}
where Eq.~\eref{i2} is the standard model prediction given in
Ref.~\cite{Gro00}. Note, that these two values are in agreement with
each other. Recently there has been some discussion of the accuracy of the
theory for Cs; in Ref.~\cite{BW99} the accuracy of the theoretical
amplitude in Cs was estimated to be 0.4\%, but at present there seems to
be a consensus that it is closer to 1\%.

In this paper, we report a new calculation of the PNC $E1$-amplitude
for the $6p_{1/2} \rtw 6p_{3/2}$ transition in thallium. Combining this
new calculation with the most accurate experiment \cite{VMM95}, leads to
a value of the weak charge of the $^{205}$Tl nucleus that can also
be compared with the standard model prediction \cite{Gro00}:
\begin{eqnarray}
        &&Q_W(^{205}{\rm Tl})=-113\,(1)_{\rm expt}(3)_{\rm theor},
\label{i3} \\
        &&Q_W^{\rm SM}(^{205}{\rm Tl})=-116.7\,(1).
\label{i4}
\end{eqnarray}
Again, these two values are in agreement with each other. The atomic
theory involved in both calculations is similar, as are the possible
sources of theoretical uncertainty. That can possibly explain why
the central values in Eq.~\eref{i1} and Eq.~\eref{i3} are shifted in the
same direction from the standard model predictions (note that in
both cases the theoretical uncertainty dominates the
experimental one).

The structure of this paper is as follows: In Sec.~\ref{calc} we
briefly describe the method of the calculation. More details can be
found in Refs.~\cite{DFK96b,DKPF98,KP99}. Results of our calculation and
a discussion of its accuracy are given in Sec.~\ref{disc}. Our
conclusions are summarized in Sec.~\ref{conc}.

\section{Calculational details}
\label{calc}

One important difference between thallium and cesium is that the former
atom has three valence electrons above a relatively compact and rigid
core, while the latter has only one. Correlations between three
valence electrons are not small and can not be treated accurately
by many-body perturbation theory (MBPT). In Ref.~\cite{DFK96b}, it was
suggested that MBPT be combined with the configuration interaction
(CI) method for such atoms. The latter method is well suited to
account for correlations between a few
valence electrons, while the former method allows one to treat
core-valence and core-core correlations. This combined
CI+MBPT method was first
tested in calculations of energy spectra of Tl \cite{DFK96b},
then for Ca, Sr, Ba, and Yb \cite{KP97}. Later it was extended to the
calculations of observable such as hyperfine structure constants
\cite{DKPF98} and polarizabilities \cite{KP99}.

There is a significant difference in MBPT for one-electron
atoms, such as Cs, and three-electron atoms such as Tl. For
one-electron atoms one has to calculate only MBPT diagrams with
one external line, while for many-electron atoms there are also
diagrams with two, three, and more external lines. For
combinatorial reasons the number of such diagrams grows rapidly
with the number of external lines, making calculations
for many-electron atoms much
more complicated. Fortunately, the three-particle diagrams
appear to be small for Tl \cite{DFK96b}. If the
three-particle diagrams are neglected, the effective Hamiltonian
for valence electrons is a two-particle operator, which not only
drastically reduces the number of diagrams at the MBPT stage of the
calculation, but is also essential for the CI stage of the
calculation because the Hamiltonian matrix remains sparse.

It was recently shown for Cs \cite{Der00,DHJS01,KPT01} that the Breit
interaction correction is larger than earlier estimated
 \cite{BSJ92}. For this reason we have included the dominant
magnetic part of the Breit interaction (the so-called Gaunt
interaction) at all stages of the present calculation. The Dirac-Fock
equations were solved in the Coulomb-Gaunt approximation for the
ground state of the Tl$^+$ ion (the $V^{N-1}$ approximation) with
the help of the code described in Ref.~\cite{BDT77}.

We used different basis sets for the CI and MBPT parts of the
calculation. That enabled us to improve the CI convergence without
having an enormous Hamiltonian matrix. The core and valence orbitals
were taken from the numerical solution of the Dirac-Fock equations
on a radial grid. The CI basis set included virtual orbitals formed
using the method suggested in Ref.~\cite{Bv83}, while virtual states
for the MBPT basis set were formed from $B$-splines \cite{SJ96}.
The Dirac-Fock Coulomb-Gaunt Hamiltonian was diagonalized on both
sets. Two different variants of the MBPT basis set were used for
calculating diagrams for the effective Hamiltonian and for solving
the RPA equations. The former included partial waves with $l=5$,
while the latter included more orbitals but was restricted to $l\leq 2$.

We use the Brillouin-Wigner variant of MBPT. In this
formalism, the effective Hamiltonian for the valence electrons is
symmetric but energy-dependent. It was shown earlier \cite{KP99tr}
that the accuracy of the theory can be improved by calculating the
Hamiltonian at shifted energies. The optimal valence energy for Tl
was found to be $-1.64$~a.u..

For some observables such as polarizabilities, Stark-induced
amplitudes and PNC amplitudes, one needs to sum over
intermediate states or solve the corresponding inhomogeneous equations
(Sternheimer \cite{Ste50} or Dalgarno-Lewis \cite{DL55} method).
Here we apply the Sternheimer-Dalgarno-Lewis method
to the valence part of the problem as described
in~\cite{KPF96,KP99}.

\section{Results and discussion}
\label{disc}

\subsection{Test Calculations of the Spectrum and Observables.}

\begin{table}
\caption{Spectrum of Tl for Coulomb and Coulomb-Gaunt potentials.
The ground state three-electron binding energy $|E_{\rm val}|$
corresponds to the sum of the first three ionization potentials.
For other levels, we give transition frequencies $\Delta$. The effective
Hamiltonian for valence electrons is calculated for the energy
$E_{\rm val}=-1.64$~a.u.
\label{t_tl_e}}

\begin{tabular}{lcdddc}
\hline
\hline
&& \multicolumn{2}{c}{CI+MBPT}
& \multicolumn{2}{c}{Experiment}
\\
&& \multicolumn{1}{c}{C}
& \multicolumn{1}{c}{C-G}
\\
\hline
$6p_{1/2}$ & $|E_{\rm val}|$ (a.u.)& 2.0742 & 2.0720 & 2.0722 &\cite{RS85}
\\
\hline
$6p_{3/2}$ &               &   7925 &   7836 &   7793 &  \\
$7s_{1/2}$ &               &  26583 &  26455 &  26478 &   \\
$7p_{1/2}$ &$\Delta$ (1/cm)&  34193 &  34087 &  34160 &\cite{Moo58}\\
$7p_{3/2}$ &               &  35215 &  35098 &  35161 &   \\
$6d_{3/2}$ &               &  36363 &  36208 &  36118 &   \\
$6d_{5/2}$ &               &  36469 &  36321 &  36200 &   \\
\hline
\hline
\end{tabular}
\end{table}
\begin{table}
\caption{Hyperfine constants for $^{205}$Tl (MHz). First row gives
Dirac-Fock values and the following rows give various corrections
described in the text.
\label{t_tl_hfs}}
\begin{tabular}{lrrrrrrr}
\hline \hline
& \multicolumn{1}{c}{$A_{6p_{1/2}}$}
& \multicolumn{1}{c}{$A_{6p_{3/2}}$}
& \multicolumn{1}{c}{$A_{7s_{1/2}}$}
& \multicolumn{1}{c}{$A_{7p_{1/2}}$}
& \multicolumn{1}{c}{$A_{7p_{3/2}}$}
& \multicolumn{1}{c}{$A_{6d_{3/2}}$}
& \multicolumn{1}{c}{$A_{6d_{5/2}}$}
\\
\hline
DF           & 17339 &  1291 &  7579 & 1940 & 187 &   21 &    9
\\
CI           &   924 &$-1369$&  3799 &$-102$& 112 &$-185$&  391
\\
$H_{\rm eff}$& 3428 & $-45$& 765 & 331 &$-56$& 114 &$-226$
\\
$A_{\rm RPA}$& 959  & 359  & 1031 & 103 & 73 & 5 & 15
\\
$A_{\sigma} $&$-1071$&  $-31$& $-269$& $-92$&$ -9$&    3 &  $-5$
\\
$A_{\rm sbt}$&$-1389$& $-161$& $-75$&$-113$&$-19$&$ -19$&$ -8$
\\
$A_{\rm tp} $& 1731 & 120 & $-22$& 133 & 4 & 21 & 7
\\
$A_{\rm SR}$ & 209 & 88 & $-29$& 14 & 6 &$ -1$& 0
\\
Norm.        &$ -467$&$   -4$&$ -113$&$ -20$&$ -3$&$   0$&$   0$
\\
\hline
Total        & 21663 &   248 & 12666 & 2193 & 295 & $-41$ &  183
\\
Theor.\footnotemark[1]
 &{21760} &{$-$1919} &{12470} &{ 2070} &{  195} &        &       \\
Theor.\footnotemark[2]
 & {21300}&{  339}&{12760}&\\
Theor.\footnotemark[3]
 &{21623} &{  264} &{12307} &{ 2157} &{  315} &{$-$35} &{  184}\\
Expt.       & 21311 &   265 & 12297 & 2155 & 309   & $-43$ &  229
\\
\hline \hline
\end{tabular}

\footnotemark[1] Ref.\cite{DFSS87}
\footnotemark[2] Ref.\cite{MP95}
\footnotemark[3] Ref.\cite{DKPF98}
\end{table}

To control the accuracy of our calculation of the PNC
amplitude, we calculated as many different observables as possible.
These results are presented in Tables~\ref{t_tl_e}
--~\ref{t_tl_al}. There are some differences with earlier
calculations \cite{DFK96b,DKPF98}. Firstly, we use the Coulomb-Gaunt
potential here instead of the Coulomb potential used in those earlier
calculations. In Table~\ref{t_tl_e}, we compare our theoretical
Coulomb-Gaunt energies with the measured spectrum. We also give Coulomb
energies for comparison. One can see that the Gaunt correction to the
fine structure is rather large, especially for the ground state.
Both splittings $6p_{1/2}$--$6p_{3/2}$ and $7p_{1/2}$--$7p_{3/2}$
are significantly improved when the Gaunt interaction is included. The
overall agreement with experiment is also improved, though not
as dramatically.

Tables~\ref{t_tl_hfs} and \ref{t_tl_e1} present results of our
calculations of hyperfine structure constants and
$E1$-amplitudes. In Table~\ref{t_tl_hfs}, we list all
corrections to the initial Dirac-Fock values of hyperfine constants.
The CI correction
accounts for mixing of configurations. The $H_{\rm eff}$
correction corresponds to the solution of the CI equations with the
effective Hamiltonian and accounts for core polarization
effects. The remaining corrections correspond to the different
terms in the effective operator of the hyperfine interaction
\cite{DKPF98}. In this calculation we included the structural
radiation ($A_{SR}$) correction \cite{DFSS87a}, which was omitted in
\cite{DKPF98}. The most important contributions are associated with
the RPA and the Brueckner ($A_{\sigma}$) corrections to the
effective operator. Two other large contributions from subtraction
($A_{\rm sbt}$)
and two-particle corrections ($A_{\rm tp}$) almost
cancel each other.

One can see that the Dirac-Fock approximation for the hyperfine
constants works reasonably well only for $6p_{1/2}$ and
$7p_{1/2}$ levels. For some of the other levels, even the sign and
order of magnitude of the constants are wrong. That affects the
final accuracy of the theory, which is better than 2\% for the
$np_{1/2}$ levels, about 3\% for the $7s$ level, and worse for
other small constants. It is actually quite surprising, that the
CI+MBPT method gives rather accurate values even when the
Dirac-Fock approximation fails completely.

\begin{table}
\caption{Reduced matrix elements for $E1$-amplitudes in $L$-gauge (a.u.).}
\label{t_tl_e1}

\begin{tabular}{rlddddl}
\hline \hline
&
& \multicolumn{1}{c}{DF}
& \multicolumn{1}{c}{CI}
& \multicolumn{1}{c}{Total}
& \multicolumn{1}{r}{Theor.}
& \multicolumn{1}{c}{Expt.}
\\
\hline
$6p_{1/2}\rtw$ &$7s_{1/2}$
&  2.049 &  1.863 &  1.77& 1.72\footnotemark[1]& 1.81(2)\footnotemark[3] \\
&&       &        &      & 1.78\footnotemark[2]&         \\
 $\rtw$ &$6d_{3/2}$
&  2.722 &  2.454 &  2.30& 2.39\footnotemark[1]& 2.30(9)\footnotemark[3] \\
$6p_{3/2}\rtw$ &$7s_{1/2}$
&  3.966 &  3.466 &  3.35& 3.18\footnotemark[1]& 3.28(4)\footnotemark[3] \\
&&       &        &      & 3.31\footnotemark[2]&         \\
 $\rtw$ &$6d_{3/2}$
&  1.633 &  1.472 &  1.40& 1.39\footnotemark[1]& 1.38(7)\footnotemark[3] \\
 $\rtw$ &$6d_{5/2}$
&  4.840 &  4.292 &  4.08&                     & 4.0(2)\footnotemark[3]  \\
&&&      &        &                            & 3.8(2)\footnotemark[4]  \\
$7p_{1/2}\rtw$ &$7s_{1/2}$
&  6.618 &  6.152 &  5.96&                     & 5.87(8)\footnotemark[5] \\
 $\rtw$ &$6d_{3/2}$
& 11.980 & 10.874 & 10.86&   \\
$7p_{3/2}\rtw$ &$7s_{1/2}$
&  8.794 &  8.252 &  7.98&                    & 7.88(11)\footnotemark[5] \\
 $\rtw$ &$6d_{3/2}$
&  5.395 &  4.887 &  4.90&   \\
 $\rtw$ &$6d_{5/2}$
& 16.300 & 14.799 & 14.88&   \\
\hline \hline
\end{tabular}

\footnotemark[1]~Ref.\cite{DFSS87a}
\footnotemark[2]~Ref.\cite{MP95}
\footnotemark[3]~Refs.\cite{HB72,GL64}
\footnotemark[4]~Ref.\cite{PS63}
\footnotemark[5]~Ref.\cite{JWD86}

\end{table}

The MBPT corrections to the $E1$ amplitudes in the length gauge are
much smaller than for the hyperfine constants and the final values
are closer to the initial Dirac-Fock ones. Usually, the $V$-form of
the $E1$ amplitude is less accurate for the optical transitions in
neutral atoms. Therefore, we used the $L$-form for the calculations
presented in Table \ref{t_tl_e1}. Accurate experimental results
are available only for the $np_j \rtw 7s$ amplitudes. For these four
amplitudes the difference between theory and experiment is
within 2\%. Taking into account the smallness of the MBPT corrections
to the $E1$ amplitudes, we assume this to be the characteristic accuracy
of the theory for $E1$ amplitudes.

\begin{table}
\caption{Polarizabilities of the $6p_j$-levels of Tl in a.u.. Atomic
polarizability includes valence and core contributions. The column
$\delta$Core accounts for the change of the core polarization due
to the fact that some valence orbitals are occupied.}
\label{t_tl_al}

\begin{tabular}{lrrrrr}
\hline \hline
& \multicolumn{1}{c}{Valence}
& \multicolumn{1}{c}{Core}
& \multicolumn{1}{c}{$\delta$Core}
& \multicolumn{1}{c}{Total}
& \multicolumn{1}{c}{Expt. \cite{Gou76}}
\\
\hline
$\alpha_0(6p_{1/2})$ &   43.47 &  6.23 &$-0.51$&  49.2  &   \\
$\alpha_0(6p_{3/2})$ &   73.79 &  6.23 &$-0.48$&  79.6  &   \\
$\alpha_2(6p_{3/2})$ & $-25.04$&  0    &$ 0.06$&$-25.0$ &$-24.2(3)$   \\
\hline \hline
\end{tabular}
\end{table}

As an additional test of the theory, we calculated the tensor
polarizability of the $6p_{3/2}$ level, which was accurately
measured by Gould \cite{Gou76} (see Table~\ref{t_tl_al}). The
contribution to the polarizability was calculated by solving the
inhomogeneous equation for the effective operators in the valence
space. The core contribution was calculated as a direct sum of the
RPA amplitudes over Dirac-Fock virtual orbitals. When calculating
the polarizability of the core of the neutral atom, we should
exclude occupied states from the sum over the intermediate
states. The corresponding correction is given in the column
$\delta$Core of Table \ref{t_tl_al}. The 3.5\% difference between
theory and experiment for $\alpha_2(6p_{3/2})$ is in
agreement with our estimate of the theoretical accuracy of the
$E1$ amplitudes.

We have omitted all three particle diagrams in the
effective Hamiltonian. These diagrams are strongly suppressed by
the small overlap between valence and core orbitals \cite{DFK96b}.
We calculated these diagrams for the small number of leading
configurations and determined corrections to the energies and to
the hyperfine constants. The valence energies changed by $\sim
10^{-4}$~a.u.\ and corrections to the hyperfine constants were on
the order of a few MHz, confirming that three particle
corrections for Tl are well below the accuracy of the present calculation.
We, therefore, neglected three-particle diagrams
in all other calculations.

\subsection{PNC Amplitude.}
\label{pnc}

The PNC interaction of the electron with the weak charge of the
nucleus has the form:
\begin{eqnarray}
        H_P &=& -\frac{G_{\rm F}}{2 \sqrt{2}}Q_W \gamma_5 \rho(\bm r),
\label{pnc1}
\end{eqnarray}
where $G_{\rm F}=2.2225\cdot 10^{-14}$~a.u.\ is the Fermi constant
of the weak interaction, $\gamma_5$ is the Dirac matrix, and
$\rho(\bm r)$ is the neutron density of the nucleus. Calculation of
the PNC amplitude is similar to calculation of polarizabilities.
Firstly, we must determine the effective operator $H_{P,\rm eff}$ for
the PNC interaction of the valence electrons with the nucleus in
the same way that we determined effective operators for the hyperfine
interaction and the electric dipole moment operators. We must then
solve the RPA equations for the operator \eref{pnc1},
calculate the Brueckner $H_{P,\sigma}$, subtraction $H_{P,\rm sbt}$,
two-particle $H_{P,\rm tp}$, and structural radiation $H_{P,\rm SR}$
corrections.

\begin{table}
\caption{$E1_{\rm PNC}$ amplitude for $6p_{1/2} \rtw 6p_{3/2}$
transition in $^{205}$Tl in the units ${\rm i}
\cdot 10^{-11}(-Q_w/N)$~a.u.
\label{t_tl_pnc} }
\begin{tabular}{ldd}
\hline \hline
right hand side:& \multicolumn{1}{c}{$H_P|6p_{1/2}\rangle$}
& \multicolumn{1}{c}{$H_P|6p_{3/2}\rangle$}\\
\hline
CI & -34.20 & -29.88  \\
$H_{\rm eff}$ \& RPA & -7.26 & +0.01  \\
Brueckner            & +1.29 & +1.12  \\
Subtraction          & +1.03 & +0.77  \\
Two-particle         & -0.29 & -0.53  \\
Struc.\ Radiation    & -0.04 & -0.02  \\
Core sum             & -0.16 & +0.06  \\
Subtotal             &
\multicolumn{2}{d}{-68.1} \\
Normalization        &
\multicolumn{2}{d}{+1.4} \\
Total                & \multicolumn{2}{d}{-66.7}
\\
\hline \hline
\end{tabular}
\end{table}

Once the effective operators are formed, we solve two
inhomogeneous equations:
\begin{eqnarray}
        (E_{6p_{1/2}} - H_{\rm eff}) \Psi_{6p_{1/2},m}^{(P)}
        &=& H_{P,\rm eff}|\Psi_{6p_{1/2},m}\rangle,
\label{pnc2}\\
        (E_{6p_{3/2}} - H_{\rm eff}) \Psi_{6p_{3/2},m}^{(P)}
        &=& H_{P,\rm eff}|\Psi_{6p_{3/2},m}\rangle.
\label{pnc3}
\end{eqnarray}
Afterward, the PNC amplitude $E1_{\rm PNC}$ for the $6p_{1/2}
\rtw 6p_{3/2}$ transition can be readily calculated with the help of the
(effective) electric dipole moment operator $\bm{D}= -e\bm{r}$:
\begin{eqnarray}
        E1_{\rm PNC}
        &=& \langle \Psi_{6p_{3/2}}||
        D_{\rm eff}||\Psi_{6p_{1/2}}^{(P)}\rangle
\nonumber \\
        &+&
        \langle \Psi_{6p_{3/2}}^{(P)}||
        D_{\rm eff}||\Psi_{6p_{1/2}}\rangle.
\label{pnc4}
 \end{eqnarray}

Alternatively, instead of \eref{pnc2} and \eref{pnc3} we can solve
inhomogeneous equations with the operator $D_{\rm eff,0}$ on the
right hand side. The reduced matrix element of the PNC
amplitude is then given by:
\begin{eqnarray}
        &&E1_{\text{PNC}}
        = (-1)^{\frac{3}{2}-m}
        \left(\begin{array}{ccc}
        \frac{3}{2} & 1 & \frac{1}{2} \\
        -m & 0 & m
        \end{array}\right)^{-1}
\label{pnc5}
        \\ &&\times
        \left(\langle \Psi_{6p_{3/2}}^{(D)}|
        H_{P,\rm eff}|\Psi_{6p_{1/2}}\rangle
        +
        \langle \Psi_{6p_{3/2}}|
        H_{P,\rm eff}|\Psi_{6p_{1/2}}^{(D)}\rangle\right),
\nonumber
\end{eqnarray}
Equations~\eref{pnc4} and \eref{pnc5} should give identical results;
a comparison of results obtained using the two equivalent methods
can be used to check numerical accuracy of the calculations. On the
other hand, it is easier to calculate the two-particle part of the
operator $H_{P,\rm eff}$ using Eq.~\eref{pnc5} and the two-particle
part of the operator $D_{\rm eff}$ using Eq.~\eref{pnc4}. The
results of the calculation based on Eqs.~\eref{pnc2}
---~\eref{pnc4} are given in Table~\ref{t_tl_pnc}. The
inhomogeneous equations \eref{pnc2} and \eref{pnc3} do not account
for the sum over intermediate core states. This sum was calculated
explicitly in the RPA approximation and the corresponding results are
also given in Table~\ref{t_tl_pnc}.

\paragraph{Accuracy of the calculation.}
The principal theoretical uncertainty in calculations for atoms such as
Tl is associated with higher orders in the residual two-electron
interaction. The effective operator method accounts for some
important higher-order MBPT corrections. The valence-valence
correlations are considered non-perturbatively within the CI method. In
addition, by diagonalizing the effective Hamiltonian, we account for
Brueckner and screening correction to all orders. The RPA equations
for the PNC and the electric-dipole interactions effectively sum
infinite chains of diagrams. However, already in third-order
MBPT, there are diagrams that are neglected here.

It is known that for neutral atoms, the residual two-electron
interaction is not small and it is very difficult to estimate the
accuracy of different approximate methods. Therefore, we made a
number of test calculations described above. We have seen that,
even for hyperfine structure constants where the MBPT corrections
are huge, the accuracy of the theory for the large constants is
still about 2 --~3\%, while for the $E1$ amplitudes the accuracy is
better than 2\%. According to Table \ref{t_tl_pnc}, the MBPT
corrections to the PNC amplitude are rather small. Therefore, we
estimate the error caused by the neglect of the higher-order terms in
the residual interaction to be about 2\%.

The other possible sources of errors are QED corrections and
the nuclear size effects. Below, we discuss both of them in some
detail.

\paragraph{QED corrections to PNC amplitude.}
The most important radiative corrections to the PNC interaction
given in Eq.~\eref{pnc1}
correspond to  very small distances and should be
calculated within the electroweak theory. These corrections lead to
scaling of the weak charge of the nucleus. The leading-order (tree-level)
value of the weak charge for $^{205}$Tl is:
\begin{eqnarray}
        Q_W(0) &=& -N + Z(1-4 \sin^2 \theta_W) = -117.9,
\label{pnc6}
\end{eqnarray}
where $N$ and $Z$ are the numbers of neutrons and protons in the
nucleus. The Weinberg angle $\theta_W$ here is taken at the energy
of $Z$-pole. Radiative corrections change $Q_W(0)$ by 1\%; the
resulting value, taken from Ref.~\cite{Gro00}, is given in
Eq.~\eref{i4}.

In addition to these radiative corrections, there are the QED radiative
corrections specific to heavy atoms \cite{LS94,Bed99}. These
corrections are dominated by vacuum-polarization effects and
for hydrogen-like ions with $Z \approx 80$ were calculated to be
0.9\% \cite{Bed99}. For the valence electron of a neutral atom,
 screening of the vacuum-polarization potential by relaxation
of the core should be taken into account. It was recently shown
that for the Breit interaction this screening significantly reduces
corrections to the valence amplitudes \cite{LMY89,DHJS01,KPT01}.
Thus, the actual radiative corrections to the PNC amplitude in Tl could
be few times smaller.

Finally, there is the Breit correction to the PNC amplitude. The
magnetic part of the Breit interaction (the Gaunt interaction) is
known to be larger than the neglected
retardation part of the Breit operator \cite{LMY89}
by a factor of three or more.
As we pointed out above, the Gaunt interaction should be included
self-consistently at all stages of calculations. It is difficult to
isolate the Gaunt correction in the self-consistent approach. To do
that, one must repeat all calculations in the Coulomb
approximation and take a difference. The calculations for Tl are
rather time consuming, so for the Coulomb case we did only the CI
calculation. At that level of approximation, the Gaunt correction to the
PNC amplitude was about $-0.5$\%.
We think that the overall Gaunt correction may be up to two times
larger. The retardation correction, which is neglected here, should
be at least few times smaller than the Gaunt correction and the
total uncertainty from QED effects should be within 1\%.

\paragraph{Nuclear size effects.}
The PNC amplitude is sensitive to both proton and neutron
distributions: the former determines electronic wave function
inside the nucleus and the latter determines the weak charge
distribution in Eq.~\eref{pnc1}. In our calculations we approximate the
nucleus $^{205}$Tl by a uniform ball of the radius $0.1334 \cdot
10^{-3}$~a.u. This value corresponds to a root-mean-square charge radius
$r_{\rm N}=5.470(5)$~fm~\cite{DDD87}.

This model assumes that (i) proton and neutron radii of the nucleus
are the same: $r_{\rm N}^{(p)}=r_{\rm N}^{(n)}=r_{\rm N}$, and (ii)
the nucleus has a sharp edge.  Generally speaking, both assumptions
are incorrect. Therefore, it is important to estimate the errors
associated with each of them. That was recently done in
Refs.~\cite{JS99,Hor01}:
\begin{eqnarray}
        \frac{\delta E1_{\rm PNC}}{E1_{\rm PNC}}
       \! \!&\approx& \!\! -0.39 \frac{\delta r_{\rm N}}{r_{\rm N}}
         -0.14 \frac{r_{\rm N}^{(n)}-r_{\rm N}^{(p)}}{r_{\rm N}}
         + 0.03 \eta,
\label{d1} \\
        \eta &=& \frac{21}{25}
        \frac{\langle r_{\rm N}^4 \rangle}
        {\langle r_{\rm N}^2 \rangle^2}-1.
\label{d2}
\end{eqnarray}
The first two terms in \eref{d1} give the dependence of the PNC
amplitude on the nuclear radius $r_{\rm N}$ and on the
difference $r_{\rm N}^{(n)}-r_{\rm N}^{(p)}$. At present there are
no accurate experimental data on neutron radii.  However,
theory predicts that for heavy nuclei $r_{\rm N}^{(n)}-r_{\rm
N}^{(p)} \approx 0.1-0.3$~fm \cite{Hor01}. The parameter $\eta$ is
defined so that it vanishes for a uniform distribution
\cite{JS99}, therefore, the last term in
\eref{d1} describes the dependence of the PNC amplitude on the
details of the nuclear distribution. For a real nucleus, $\eta$ is
about 0.1 \cite{Hor01}.

Substituting into Eq.~\eref{d1}, we find that the correction
to the PNC amplitude from the details of the nucleon distribution is
\begin{eqnarray}
        &&\frac{\delta E1_{\rm PNC}}{E1_{\rm PNC}}
        \approx-0.003\,(2).
\label{d3}
\end{eqnarray}
This correction is seen to be small in comparison with other
ones.

\section{Conclusion}
\label{conc}

Our final value for the PNC amplitude of the $6p_{1/2} \rtw
6p_{3/2}$ transition is in good agreement with the most accurate
previous calculation of Dzuba \etal \cite{DFSS87a} (the references
to  earlier calculations can be found in \cite{Khr91}):
\begin{equation}
        \frac{E1_{\rm PNC}}{\text{i} \,10^{-11}(-Q_{\rm w}/N)}
        =\left\{\begin{array}{lcl}
        -(66.7\pm 1.7) &\, &\text{this work,}\\
        -(66.1\pm 2.0) &   &\text{Dzuba \etal}
        \end{array}\right.
\label{conc1}
\end{equation}
In the experiment the following ratio is measured:
\begin{equation}
        {\cal R} \equiv
        {\rm Im} \left( \frac{E1_{\rm PNC}}{M1} \right),
\label{conc2}
\end{equation}
where $M1$ is the $6p_{1/2} \rtw 6p_{3/2}$ magnetic-dipole transition
amplitude. We have calculated this amplitude in two different ways.
Firstly, we did the CI+MBPT calculation using the effective Hamiltonian.
Secondly, we did a third-order MBPT calculation in which Tl was treated
as a one-electron atom:
\begin{equation}
        M1 = \left\{\begin{array}{lll}
        4.145 \cdot 10^{-3}\text{ a.u.} &\quad & \text{CI+MBPT-II,}\\
        4.149 \cdot 10^{-3}\text{ a.u.} &\quad & \text{MBPT-III.}\\
        \end{array}\right.
\label{conc3}
\end{equation}
These two calculations account for different correlation effects,
but give very close results. In general, the amplitudes for the
allowed $M1$-transitions are very stable and can be calculated quite
reliably. The experimental value of this amplitude follows from
the measurements of the quadrupole amplitude $E2$
\cite{Vet95} and the ratio
$\chi\equiv\omega/(2\sqrt{3} c)\,E2/M1$ \cite{MT99}:
\begin{eqnarray}
\left.
\begin{array}{lcl}
        E2 &=& 13.29\,(3)  \\
        \chi &=& 0.2387\,(40)\\
\end{array}\right\} \Rightarrow
        M1 = 4.16\,(7)\, 10^{-3},
\label{conc4}
\end{eqnarray}
which is in agreement with theoretical results of Eq.~\eref{conc3}.

Using the theoretical value for the $M1$ amplitude \eref{conc3} and
the standard model value of the weak charge \eref{i4}, we get the
following result for the PNC rate $\cal R$:
\begin{eqnarray}
        {\cal R}\left(Q_W = -116.7\right)
        &=& -15.2\,(4)\cdot 10^{-8}.
\label{conc5}
\end{eqnarray}
The experimentally measured PNC rate for the $6p_{1/2} \rtw
6p_{3/2}$ transitions is
\begin{eqnarray}
        {\cal R} = \left\{\begin{array}{lll}
        -14.68\,(17), & \text{Vetter \etal}  & \mbox{\cite{VMM95}}, \\
        -15.68\,(45), & \text{Edwards \etal} & \mbox{\cite{EPBN95}}.
        \end{array}\right.
\label{conc6}
\end{eqnarray}
These experimental results formally contradict one another. Recently,
Majumder and Tsai suggested \cite{MT99} that the discrepancy could be
due to the different values of the parameter $\chi$ used in the
analysis by two groups. They accurately measured $\chi$
\eref{conc4} and used their value to rescale the
experimental result from Ref.~\cite{EPBN95} to find
\begin{eqnarray}
        {\cal R} = -14.71\,(45),  \text{ Majumder and Tsai~\cite{MT99}}.
\label{conc7}
\end{eqnarray}
This scaled value is in agreement with the measurement
\cite{VMM95}, where a nearly identical value of $\chi$ was used, and all
three values are in agreement with the theoretical result
\eref{conc5} for the $Q_W=Q_W^{\rm SM}$.  We use the best
experimental result \cite{VMM95} together with our calculation \eref{conc5}
to derive the experimental value of the weak charge for $^{205}$Tl:
\begin{eqnarray}
        Q_{\rm W}(^{205}{\rm Tl})
        = -113\,(1)_{\rm expt}(3)_{\rm theor}\ .
\label{conc8}
\end{eqnarray}

\section*{acknowledgment}
We are grateful to L. N. Labzowsky, M. S. Safronova, U. I. Safronova, and
I. M. Savukov for helpful discussions. One of us (MK) is grateful to
the Notre Dame University for hospitality. The work of WRJ was supported in
part by the U. S. National Science Foundation under grant No.\ 99-70666.


\end{document}